%% file: main.tex
\documentclass[aps,
twocolumn, 
showpacs,
prl,
superscriptaddress,
tightenlines,
10pt,
longbibliography] 
{revtex4-2}
\usepackage{graphicx}
\usepackage{xcolor}
\usepackage{braket}
\usepackage{hyperref}
\usepackage{comment}
\usepackage{amsmath}
\usepackage{appendix}
\usepackage{physics}
\usepackage{siunitx}
\usepackage{microtype}
\usepackage{lmodern}
\usepackage{import}
\usepackage{xifthen}
\usepackage{pdfpages}
\usepackage{transparent}
\graphicspath{{figures/}}

\def\g2{g^{(2)}(0)}
\def\gi{g^{(2)}_{\mathrm{in}}(0)}
\def\go{g^{(2)}_{\mathrm{out}}(0)}

\makeatletter
\newcommand*{\defeq}{\mathrel{\rlap{%
                     \raisebox{0.3ex}{$\m@th\cdot$}}%
                     \raisebox{-0.3ex}{$\m@th\cdot$}}%
                     =}
\makeatother

\UseRawInputEncoding
\usepackage{lipsum}
\makeatletter
\AtBeginDocument{\let\LS@rot\@undefined}
\makeatother

\begin{document}
\title{Strongly nonlinear interaction between non-classical light \\ and a blockaded Rydberg atomic ensemble}

\author{Jan Lowinski}
\email{jan.lowinski@icfo.eu}
\affiliation{ICFO -- Institut de Ciencies Fotoniques, The Barcelona Institute of Science and Technology, Spain}

\author{Lukas Heller}
\affiliation{ICFO -- Institut de Ciencies Fotoniques, The Barcelona Institute of Science and Technology, Spain}

\author{F\'elix Hoffet}
\affiliation{ICFO -- Institut de Ciencies Fotoniques, The Barcelona Institute of Science and Technology, Spain}

\author{Auxiliadora Padr\'on-Brito}
\affiliation{ICFO -- Institut de Ciencies Fotoniques, The Barcelona Institute of Science and Technology, Spain}

\author{Klara Theophilo}
\altaffiliation[Current address: ]{National Quantum Computing Centre, OX11 0QX Didcot, United Kingdom}
\affiliation{ICFO -- Institut de Ciencies Fotoniques, The Barcelona Institute of Science and Technology, Spain}

\author{Hugues de Riedmatten}
\email{hugues.deriedmatten@icfo.eu}
\affiliation{ICFO -- Institut de Ciencies Fotoniques, The Barcelona Institute of Science and Technology, Spain}
\affiliation{ICREA -- Instituci\'o Catalana de Recerca i Estudis Avan\c cats, 08015 Barcelona, Spain}

\begin{abstract}
We investigate the interaction between non-classical light with a tunable multiphoton component and a highly nonlinear medium based on cold Rydberg atoms. 
The non-classical field emitted by a DLCZ quantum memory is stored using Rydberg electromagnetically induced transparency, experiencing strong nonlinear response due to the dipole blockade.
We show that the storage efficiency in the Rydberg ensemble decreases as function of the multiphoton strength of the input field, as a result of the nonlinearity.
We also show that the autocorrelation function $\g2$ of the retrieved field after storage in the Rydberg state is considerably reduced, leading to the first demonstration of single photon filtering with non-classical input light.
Finally, we develop a simple simulation that allows us to model the effect of our medium on the input state.
This work is a step towards matter-mediated photon-photon interactions with non-classical light. 
\end{abstract}

\maketitle


Systems displaying strong nonlinear response at the few-photon level have recently drawn attention as they offer opportunities to engineer interactions between single photons \cite{Chang2014}.
In this context, several platforms have been investigated, such as single atoms coupled to optical resonators \cite{Birnbaum2005,Dayan2008,Tiecke2014,Hacker2016}, atomic ensembles \cite{Pritchard2010,Dudin2012,Parigi2012,Peyronel2012,Maxwell2013,Thompson2017,Tiarks2016}, atomic chains \cite{Prasad2020} and quantum dots in nanophotonic cavities \cite{Fushman2008,Volz2012,Kim2013,Javadi2015,deSantis2017}.
Atomic ensembles emerge as a particularly promising platform, as they allow for strong light-matter interactions without a high-finesse cavity.
In this system, few-photon nonlinearity can be achieved with electromagnetically-induced transparency to a Rydberg state if the ensemble is of a small size \cite{Peyronel2012,Paredes-Barato2014}.
The nonlinearity results from strong dipole-dipole interactions between atoms excited to the Rydberg state, which prevent multiple excitations in the ensemble.
This effect, known as the Rydberg blockade, has been widely used for proof-of-principle demonstrations such as logical gates \cite{Tiarks2019,Stolz2022}, single-photon level switches and transistors \cite{Tiarks2014,Baur2014,Gorniaczyk2014}, fast qubit detection \cite{Xu2021, Vaneecloo2022}, single-photon generation \cite{Ornelas-Huerta2020,Yang2022} and entangled-photon-pair generation \cite{Sun2022,Ye2022}.

The promise of Rydberg blockaded ensembles for quantum information processing purposes is clear.
However, so far, all proof-of-concept demonstrations with blockaded ensembles have used classical weak coherent states (WCS) as inputs, although there are records of single photons being stored in Rydberg ensembles with weaker nonlinearity \cite{Distante2017, Yu2020}. 
Nevertheless, for applications in quantum networks it is crucial to demonstrate that single photons can interact with a blockaded ensemble with single-photon non-linearity. For example, a photon-photon gate between two single photons that are part of an entangled state would allow deterministic Bell state measurements and entanglement swapping, important capabilities for scaling up quantum networks.
Using single photons as inputs have also been predicted to improve the contrast of single photon transistors \cite{Gorniaczyk2014}. 

As a step towards these applications, we report the first experimental demonstration of the interaction and storage of a correlated single photon in a highly nonlinear medium based on cold Rydberg atoms.  
We use the DLCZ protocol \cite{Duan2001} in a cold-atomic ensemble to generate heralded non-classical states of light with a tunable multiphoton component.
Those photons are then guided to another ensemble and stored in a highly excited Rydberg state using dynamical electromagnetically induced transparency (rEIT)  \cite{Fleischhauer2000, Petrosyan2011}.
We assert the single-photon-level nonlinearity of our system by comparing autocorrelation functions $\g2$ of the input and output photons, showing the first realization of single photon filtering with non-classical input states.
Additionally, we demonstrate that the nonlinearity depends only on the input Fock-state distribution of the optical field.

\begin{figure*}[t]
    \begin{minipage}[c]{0.99\linewidth}
   \def\svgwidth{\columnwidth}
   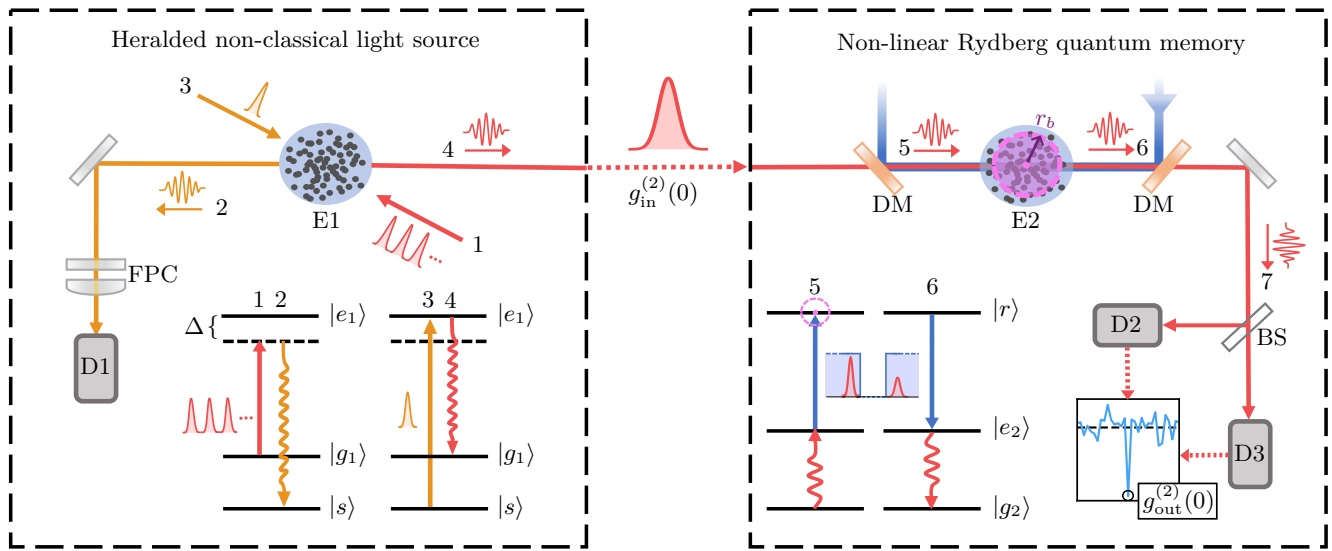
    \end{minipage}
    \caption{Schematic view of the experiment. A train of write pulses detuned by $\Delta = \SI{-40}{\mega\hertz}$ from the $\ket{g} \rightarrow \ket{e}$ transition is sent to E1 (step 1) until a Raman scattered photon is detected on detector D1 (step 2), heralding a collective excitation in E1. This spin-wave is then read out by a strong read pulse resonant with $\ket{s} \rightarrow \ket{e_{1}}$ (step 3), which produces the emission of a read photonic field in a well-defined mode (step 4). Using the nonlinearity of the Rydberg blockade effect, this field is then mapped to a single Rydberg excitation in E2 by means of Rydberg-EIT driven by a coupling beam (step 5). At a later time, the excitation is read out by reapplying the coupling beam (step 6) and the statistics of the field are studied in a HBT setup consisting of one beamsplitter and two detectors (step 7). D1(2,3) : Single-photon detector 1(2,3); E1(2): Atomic ensemble 1(2); DM: Dichroic mirror; BS: Beam-splitter; FPC: Fabry-Perot cavity; $r_{b}$: Rydberg blockade radius ($\sim \SI{13}{\micro\meter}$).}
    \label{fig:setup}
\end{figure*}


Our experimental setup consists of two ensembles of cold $^{87}$Rb atoms located in the same laboratory and connected via \SI{10}{\meter} of optical fiber cable (see \autoref{fig:setup}).
The first ensemble (E1) is used to generate heralded non-classical light using the DLCZ protocol.
The second ensemble (E2) is used as a nonlinear Rydberg EIT quantum memory with which the DLCZ light interacts. 

We start by sending a train of write pulses to E1, detuned by $\Delta = \SI{-40}{\mega\hertz}$ from the ${\ket{g_1} = \ket{5S_{1/2}, F=2, m_F=+2}}$ to ${\ket{e_1} = \ket{5P_{3/2}, F=2, m_F=+1}}$ transition.
With a low probability, this generates a write photon in the heralding mode which, upon detection at detector D1, heralds a collective spin excitation (spin wave) in the spin state $\ket{s} = \ket{5S_{1/2}, F=1, m_F=0}$.
To reduce the heralding noise, we filter unwanted frequencies with a Fabry-Perot cavity in the write mode.
In total, the transmission of a write photon in our setup (including its detection) is \SI{21}{\percent}. 
After \SI{1.6}{\micro\second}, we then send a read pulse resonant with the $\ket{s} \rightarrow \ket{e}$ transition that maps the collective excitation into a read photon.
This photon is resonant with the $\ket{g} \rightarrow \ket{e}$ transition and is emitted into a well-defined mode that depends on the phase matching conditions of the process.
It is then collected in an optical fiber and sent to the nonlinear Rydberg medium. 
The atomic parametric interaction used in the DLCZ process creates photon pairs in a two-mode squeezed state \cite{Duan2001, Hammerer2010}. 
By changing the write pulse intensity, we vary the probability of creating ($p$) and detecting a write photon ($p_w$), and therefore we can tune the multiphoton probability in the read field. 

The read photons are then frequency-shifted by an acousto-optical modulator (AOM) to match the ${\ket{g_2} = \ket{5S_{1/2}, F=2} \rightarrow \ket{e_2} = \ket{5P_{3/2}, F=3}}$ transition in E2.
They are directed to the Rydberg medium, where they propagate under EIT conditions.
The EIT coupling field is counter-propagating and resonant with the ${\ket{e_2} \rightarrow \ket{r} = \ket{90S_{1/2}}}$ transition.
The photons propagate as Rydberg polaritons strongly interacting through van der Waals interactions, which prevent multiple Rydberg excitations in the cloud \cite{Lukin2001}.
Since only one excitation can exist in the cloud at a time, only one photon can be retrieved.
As a consequence, the medium response is nonlinear and the statistics of the photonic pulses are affected.
Additionally, when the Rydberg polariton is propagating in the cloud, we can switch off the coupling field, freezing the polariton's propagation and effectively performing storage.
This is known to enhance the nonlinearity \cite{Distante2016,Padron-Brito2021a}.
Finally, the retrieved photons are detected using single-photon detectors (D2 and D3) and their statistics are measured in a Hanbury Brown-Twiss (HBT) setup.

For the DLCZ source to generate photons efficiently, its optical depth ($\mathrm{OD}$) must be large \cite{Simon2007} and the coherence time of the collective spin excitation must be longer than the time between the write and the read pulse.
We achieve both using standard techniques of magneto-optical trapping (MOT) assisted by a single retro-reflected beam dipole trap at $\SI{797}{\nano\meter}$.
We obtain a cloud with $\mathrm{OD} = 6$ and an initial temperature of $\sim\SI{80}{\micro\kelvin}$, cold enough to suppress effects of motional dephasing.
However, in each trial the ground state population is swapped back and forth between $\ket{g_{1}}$ and $\ket{s}$ (an intrinsic property of the DLCZ protocol) effectively heating up the cloud and resulting in a short dipole-trap trapping time of the order of a few ms, much shorter than the Rydberg one.

To maximize the nonlinearity of the Rydberg medium, one needs large OD in a small ensemble \cite{Firstenberg2016}.
The characteristic length is given by the dipole blockade radius $r_b$ -- radius of a sphere around an atom excited to the Rydberg level where, due to the dipole-dipole interactions, no other atom can be excited to the Rydberg state \cite{Lukin2001}.
In our case, for $n = 90$,  $r_b \approx \SI{10.5}{\micro\meter}$ \cite{Firstenberg2016, Weber2017}.
To achieve such a regime, we first trap our atoms in a MOT using similar techniques as in E1 and then transfer them into a small crossed dipole trap at \SI{852}{\nano \meter}.
In this way, we obtain a spherical cloud with $\mathrm{OD} = 11$, diameter of \SI{15}{\micro\meter} (FWHM) and temperature of $\sim\SI{40}{\micro\kelvin}$.
The EIT transparency is limited to about $\SI{60}{\percent}$ due to a large dephasing rate of the $\ket{e_r} \rightarrow \ket{r}$ transition, attributed to stray RF fields and motional dephasing.
To avoid additional dephasing we lock the lasers to a home-build reference cavity, allowing for linewidth reduction and long-term stability \cite{SM}.

Because of the different trapping cycles of the two ensembles  (\SI{12}{\milli\second} for E1 vs \SI{1.3}{\second} for E2), the overall duty cycle of the experiment is limited to $\SI{5}{\percent}$.
Besides, passive losses affect quadratically the coincidence probability in the HBT experiment. 
Both factors result in long interrogation times, which makes this experiment challenging, as high stability is required for long periods of time.


\begin{figure}[t]
    \centering
    \includegraphics[width=0.47\textwidth]{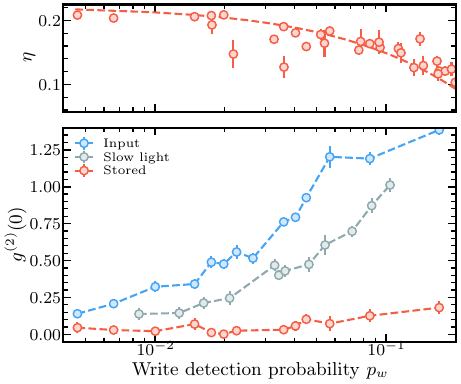}
    \caption{(top) Decay of the heralded single-photon storage and retrieval efficiency $\eta$ as a function of detected write probability $p_{w}$.
    The dashed line is a guide to the eye.
    (bottom) Autocorrelation $\g2$ of the input (blue), propagating under rEIT conditions (green), and stored (red) DLCZ read photons for different values of $p_{w}$. 
    We observe a strong reduction of the DLCZ read photons $\g2$ after their storage -- a clear manifestation of the strong nonlinearity of our Rydberg medium.
    $\go$ of pulses propagating without storage is only lowered with respect to the input due to their duration larger than the group delay (\SI{150}{\nano\s} vs. $\sim \SI{85}{\nano\s}$) and complex dynamics of propagation of such short pulses \cite{Mohl2020, Padron-Brito2021a}.
    This data was corrected for background noise induced by the blue coupling light and dark counts of the detectors (for details, see \cite{SM}).}
    \label{fig:dlcz_storage_g2}
\end{figure}

We now discuss our results. 
We start by characterizing our heralded non-classical light source.
The read photons statistics can be changed by varying the write pulse intensity during the excitation stage of the DLCZ protocol (step 1 in Fig. \ref{fig:setup}).
By increasing the write pulse intensity, and consequently increasing the probability of detecting a write photon $p_w$, we can tune the heralded autocorrelation $\gi$ of the read photons from 0.1 to 1.4, allowing us to study the response of the Rydberg medium to light with different input statistics.

The heralded DLCZ states are stored in the Rydberg medium and their storage and retrieval efficiency is measured as a function of $p_{w}$, as shown in \autoref{fig:dlcz_storage_g2}(top).
For low $p_{w}$ values, the efficiency is around \SI{20}{\percent}. 
As $p_{w}$ increases, thereby enlarging the multiphoton contribution in the heralded input field, we observe a pronounced reduction in efficiency.
This behavior is consistent with the effects of the dipole blockade, which turns higher-order Fock states into one-photon states.
Such dynamics have previously been reported for weak coherent states during slow-light propagation \cite{Peyronel2012, Bienias2020} or storage \cite{Padron-Brito2021} under rEIT conditions.
Notably, this result marks a first demonstration of non-classical light pulse storage within a strongly nonlinear Rydberg medium.

\begin{figure}
    \centering
    \includegraphics[width=0.47\textwidth]{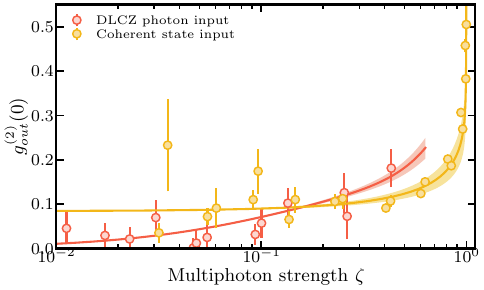}
    \caption{Autocorrelation $\go$ of DLCZ read photons (red circles) and coherent states (yellow circles) after their storage in the nonlinear medium, for different values of multi-photon strength $\zeta$ of the input state.
    The solid lines in the corresponding colors represent the Monte Carlo simulation results discussed in the main text.
    The shaded area corresponds to uncertainty of estimating losses before the storage (see \cite{SM} for details).}
    \label{fig:hard_sphere_model}
\end{figure}

We turn our attention to $\g2$, as it is a vital metric for achieving high fidelity entanglement in quantum networks \cite{Sangouard2011} (for the definition and measurement methods of $\g2$, see supplementary material \cite{SM}).
After the interaction of the DLCZ-emitted light with our nonlinear medium, we observe a pronounced suppression of the $\g2$ value, as seen in \autoref{fig:dlcz_storage_g2}(bottom), another clear indicator of the nonlinear interaction taking place.
Specifically, the $\g2$ for slow light experiences only a slight reduction, attributed to the pulse not being fully compressed within one blockade radius.
The propagation time through the medium is around \SI{85}{\nano\second} while the pulses are \SI{150}{\nano\second} long (FWHM).
It means that the nonlinearity affects only a part of a pulse at a time, what effectively decreases the strength of the nonlinearity.
The $\go$ for stored light, on the other hand, stays low for increasing value of $p_w$ and input $\gi$.
This is a clear sign of nonlinearity enhancement due to storage, as previously observed in \cite{Distante2016}.
These measurements are the first observation of a $\g2$ reduction and single photon filtering with a non-classical input for any kind of system.

Nevertheless, our $\go$ does not reach zero and rises as $p_w$ increases, implying that the ensemble might not be under full blockade.
To explain the observed trend, we develop a simple Monte Carlo simulation accounting for the influence of an imperfect blockade on different Fock states.
This model rests on the assumption that the blockade manifests as a binary effect -- a photon either becomes a polariton propagating without losses or gets scattered, referred to as the hard-sphere model \cite{Gorshkov2013, Zeuthen2017}.
The controlling parameter is the blockade radius relative to the cloud size. 
A detailed explanation can be found in the supplementary material \cite{SM}.

As we later discuss, not only do we validate our model with the collected $\go$ data for DLCZ photons, but we also study the response of our medium with input coherent states.
To facilitate the representation of the results for both data sets on a single plot, we introduce the concept of multiphoton strength $\zeta$.
For a Fock state distribution of the incoming state $p_n=\bra{n}\rho_{in}\ket{n}$, where $\rho_{in}$ is the incoming state and $\ket{n}$ is $n$-photon Fock state, the multiphoton strength $\zeta$ is defined as the probability of having two or more photons in a pulse normalized to the probability of having at least one photon:
\begin{equation}
    \zeta = \frac{\sum_{n \ge 2} p_n}{\sum_{n\ge 1} p_n}.
\end{equation}
This choice is motivated by the fact that multiphoton components adversely impact the performance of quantum networks.
In contrast, the mean photon number provides limited insights into the underlying Fock state distribution, making multiphoton strength a more informative metric.

To estimate the multiphoton strength $\zeta$ of the DLCZ read photons before entering the nonlinear medium, we measure their $\gi$ and, based on the standard assumption that DLCZ emits light in a two-mode squeezed state \cite{Duan2001}, we compute its corresponding Fock state distribution (see supplementary material for details \cite{SM}).
For WCS, we simply assume Poissonian distribution with a mean photon number backpropagated from the detected photon number.

\begin{figure}[t]
    \centering
    \includegraphics{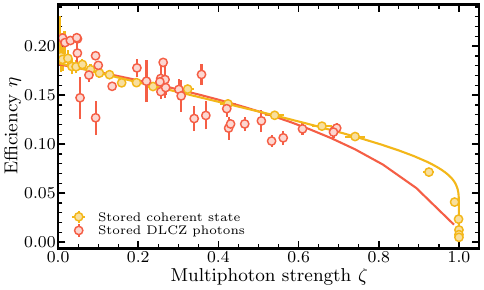}
    \caption{Rydberg memory efficiency as a function of multiphoton strength $\zeta$ for storage and retrieval of DLCZ photons (red circles) and WCS state (yellow circles). 
    The solid lines of corresponding colors show the model prediction.
    The predicted faster efficiency decay for DLCZ photons at large $\zeta$ comes from the fact that the corresponding Fock state distribution has a longer tail towards higher-order multiphoton components than for WCS.}
    \label{fig:hard_sphere_model_efficiency}
\end{figure}

The model replicates well the observed $\go$ data for both DLCZ photons and WCS, as shown in \autoref{fig:hard_sphere_model}, where we set the blockade radius to $r_b = \SI{10.5}{\micro\m}$, aligning with the value calculated for our experimental parameters \cite{Firstenberg2016}.
While the blockade radius traditionally pertains to rEIT conditions without storage, and the hard-sphere model suits very high OD regime \cite{Zeuthen2017, Bienias2020}, we attribute the agreement between the data and the simulation to storage enhancing the medium nonlinearity, akin to an OD boost \cite{Distante2016}.
More details on this can be found in supplementary material \cite{SM}.

Using our simulation, we can explain the distinct $\go$ patterns observed for DLCZ photons and WCS.
In most cases, 2-photon state is converted into 1-photon state, but occasionally the imperfect blockade allows for its survival.
Higher order components get transformed into either 1- or 2-photon states, with the 2-photon conversion probability rising with $n$.
At low $\zeta$, $\go$ values primarily reflect the amplitude of 2-photon states, meaning $\go \approx 2 p_2 / (p_1)^2$.
For WCS, since the input $\gi = 1$ at any $\zeta$, post-storage $\go$ remains flat at lower $\zeta$ values, determined by the 2-photon state's survival probability.
However, as $\zeta$ grows, higher-order components emerge, leading to an increase in $\go$.
Conversely, DLCZ photons show an increasing $\gi$ trend with $\zeta$.
After storage, the trend persists, but the values diminish due to the blockade.
The stored DLCZ photon $\go$ intersects the stored WCS $\go$ when the input DLCZ photons have $\gi \approx 1$.

It's worth noting that due to the ensemble being only partially blockaded, the nonlinearity becomes sensitive to the shape of the input Fock state distribution.
This is precisely why the $\go$ of DLCZ photons raises more rapidly than that of WCS -- the tail of the DLCZ Fock state distribution is longer than of WCS at larger values of $\zeta$.

As a side-note, we mention that the cross-correlation $g_{w,r}^{(2)}(0)$ between the write and the read photons is also affected by the storage of the read photons in the Rydberg medium, however, this change is very small and is mainly due to a temporal and frequency filtering in the EIT memory rather than the nonlinearity (for a discussion and the definition of $g_{w,r}^{(2)}(0)$ see supplementary material \cite{SM}).

Considering the impact of the partial blockade on the pulse's Fock state distribution, we can re-examine the efficiency data and simulate its decay with respect to $\zeta$. 
The comparison between our simulations and the experimental data for both DLCZ photons and WCS is depicted in \autoref{fig:hard_sphere_model_efficiency} (see supplementary materials \cite{SM} for the simulation details).
While our simulation aligns well with the WCS data, there's a some discrepancy for the DLCZ data at higher $\zeta$ values.
This inconsistency likely stems from the uncertainty in determining $\zeta$ for this dataset.
It's derived from the correspondence of $\zeta$ to the measured $p_w$, and accurate calibration of transmission losses of the write photon is crucial for its reliability \footnote{
Such calibration concerns don't affect the $\go$ data, as $\zeta$ is determined from measured input $\gi$, which is independent of the linear losses' calibration of the write photon.
}.

It's worth noting that both models align with the experimental data for WCS and DLCZ photons, despite many parameters, like mean photon number, experimental rate or input $\gi$, being very different.
Given that these models focus exclusively on the input Fock state distribution, this suggests that the Rydberg medium's nonlinear response is determined by this distribution, which confirms the prevailing understanding of the Rydberg ensemble nonlinearity.


In this work, we employed a DLCZ quantum memory as a heralded source of non-classical light, examining its interplay with Rydberg nonlinearity during storage under EIT conditions in an atomic ensemble.
This constitutes the first experiment where a non-classical state is stored in such a highly nonlinear medium, a crucial prerequisite for their applications in quantum networks.
The photon's autocorrelation $\g2$ was strongly reduced due to this interaction, demonstrating single photon filtering with quantum input light.
With a simple simulation, we explained the role of partial blockade of the ensemble in this process.
By comparing the results with quantum light and weak coherent states, we showed that the input Fock state distribution dictates the response of the medium, which aligns with established understandings of such systems.
Our results show a proof of principle that correlated single photons can be stored in a Rydberg medium with single photon level nonlinearity.
This represents a step towards the realization of photon-photon gates with true single photons.

\begin{acknowledgments}
This project received funding from the Government of Spain (PID2019-106850RB-I00 project funded by MCIN/ AEI /10.13039/501100011033; Severo Ochoa CEX2019-000910-S), from MCIN with funding from European Union NextGenerationEU (PRTR-C17.I1), from the European Union's Horizon 2020 research and innovation program under Grant Agreement No. 899275 (DAALI), from the Gordon and Betty Moore Foundation through Grant No. GBMF7446 to H. d. R, from Fundaci{\'o} Cellex, Fundaci{\'o}  Mir-Puig and from Generalitat de Catalunya (CERCA, AGAUR). 
L.H. and J.L. acknowledge funding from the European Union’s Horizon 2020 research and innovation program under the Marie Sk\l{}odowska-Curie grant agreement No. 713729.
\end{acknowledgments}

L.H. and J.L. contributed equally to this work.
\bibliography{purification}

\end{document}


\title{Supplementary Material for ``Strongly non-linear interaction between non-classical light and a blockaded Rydberg atomic ensemble"}

\author{Jan Lowinski}
\affiliation{ICFO -- Institut de Ciencies Fotoniques, The Barcelona Institute of Science and Technology, Spain}

\author{Lukas Heller}
\affiliation{ICFO -- Institut de Ciencies Fotoniques, The Barcelona Institute of Science and Technology, Spain}

\author{F\'elix Hoffet}
\affiliation{ICFO -- Institut de Ciencies Fotoniques, The Barcelona Institute of Science and Technology, Spain}

\author{Auxiliadora Padr\'on-Brito}
\affiliation{ICFO -- Institut de Ciencies Fotoniques, The Barcelona Institute of Science and Technology, Spain}

\author{Klara Theophilo}
\altaffiliation[Current address: ]{National Quantum Computing Centre, OX11 0QX Didcot, United Kingdom}
\affiliation{ICFO -- Institut de Ciencies Fotoniques, The Barcelona Institute of Science and Technology, Spain}

\author{Hugues de Riedmatten}
\affiliation{ICFO -- Institut de Ciencies Fotoniques, The Barcelona Institute of Science and Technology, Spain}
\affiliation{ICREA -- Instituci\'o Catalana de Recerca i Estudis Avan\c cats, 08015 Barcelona, Spain}

\maketitle

\section{Experimental details}

\begin{figure}[t]
    \centering
    \includegraphics[scale=1.1]{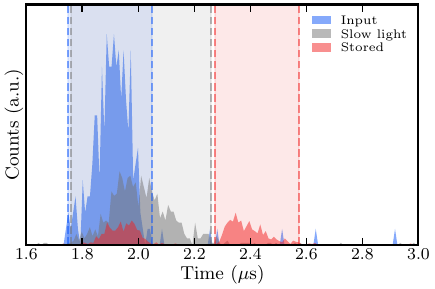}
    \caption{Typical histograms counts of the single photons before the memory (blue), after slow-light propagation (gray) and storage (red). The temporal windows in which statistics are evaluated are represented by the dotted lines. The window length for the input and the stored photon is \SI{300}{\nano\second} long, and \SI{500}{\nano\second} long for the slow-light. }
    \label{histo}
\end{figure}

\subsection{Heralded photon source}

At ensemble E1, $^{87}$Rubidium atoms are trapped from background-gas pressure in a magneto-optical trap (MOT) during \SI{8}{\milli\second}.
Subsequently, the cloud is cooled by polarization-gradient cooling (PG cooling) for \SI{500}{\micro\second}, while ramping down the magnetic field.
Atoms are then prepared in the ground state ${\ket{g_1}}$ by optical pumping, achieving an initial optical depth of $\mathrm{OD} = 5$ on the ${\ket{g_1}} \rightarrow \ket{e_1}$ transition.
To suppress atom loss by diffusion and fall in gravity, the atoms are loaded into an optical dipole trap which intersects with the ensemble.
The dipole trap consists of a focused light beam of linear polarization, at \SI{796.5}{\nano\meter}.
\SI{250}{\milli\watt} of power are focused down to a beam waist of \SI{110}{\micro\meter}.
The light is retro-reflected, forming a one-dimensional pancake lattice with potential depth of $U_0 \approx \SI{400}{\micro\kelvin}$.
The beam is incident at an angle of $\sim \ang{10}$ with the photon mode.
Then, the write photon generation starts by sending a train of write pulses.
An unsuccessful attempt is followed by a cleaning pulse, pumping the atoms back to $\ket{g_1}$.
A successful heralding is instead followed by an electronic trigger sent to ensemble E2, signaling the successful generation of a spin wave.
\SI{1.6}{\micro\second} after the write pulse, a read-out pulse retrieves the excitation as a photon.
The $1/e^2$ beam radius is \SI{69}{\micro\meter} for the photon mode and \SI{180}{\micro\meter} for the excitation pulses.
All the transitions used in photon generation belong to the D2 line at around \SI{780}{\nano\meter}.
After \SI{4}{\milli\second} of interrogation time, the magnetic field is switched on again and the sequence repeats.
The total cycle lasts $\sim \SI{12}{\milli\second}$.
The generated read photon is guided to E2, the non-linear medium.
The frequency of the photon is, however, not compatible with the transitions used in the Rydberg medium, so it is shifted by +\SI{266}{\mega\hertz} with an acousto-optic modulator to make it resonant to the $\ket{g_2} \rightarrow \ket{e_2}$ transition.

\subsection{Nonlinear medium}

At E2, $^{87}$Rubidium atoms are also trapped from background-gas pressure in a MOT during \SI{1}{\second}.
Subsequently, the cloud is compressed to increase the atomic density.
To that end, the magnetic field gradient is ramped up from \SI{20}{\gauss\per\cm} to \SI{50}{\gauss\per\cm} during \SI{9}{\milli\second}.
The atoms are then cooled with PG cooling to $\approx \SI{40}{\micro\kelvin}$ for \SI{8}{\milli\second} while shutting down the magnetic field, prepared in the ground state $\ket{g_2}$, and simultaneously loaded into a crossed dipole trap.
The dipole trap consists of two tightly focused light beams of linear polarization at \SI{852}{\nano\meter}, each with a power of \SI{1}{\watt}.
Both beams have a beam waist of \SI{34}{\micro\meter}.
One beam is horizontal, in-plane with the coupling beam and photon modes, and intersects with the photon mode at an angle of \ang{22}.
The second beam is incident from the top.
This way, a spherical cloud with $\mathrm{OD} = 11$ on the $\ket{g_2} \rightarrow \ket{e_2}$ transition is obtained, with a diameter of \SI{26}{\micro\meter} and temperature of~$\sim\SI{40}{\micro\kelvin}$, as measured by time of flight.
E2 is now ready to receive photons from either the heralded source or from a weak coherent laser pulse.

In the Rydberg medium, the photons propagate under EIT conditions.
The EIT coupling field is counter-propagating and resonant with the ${\ket{e_2} \rightarrow \ket{r}}$ transition, corresponding to a wavelength of \SI{479}{\nano\meter}.
It is derived from a 479-nm frequency converted amplified diode laser with a seed at \SI{958}{\nm}.
Depending on the measurement, the coupling is either constantly on (slow light) or switched off during propagation, mapping the Rydberg polariton onto a Rydberg state (storage).
The $1/e^2$ beam radius is \SI{6.5}{\micro\meter} for the probe and \SI{13}{\micro\meter} for the coupling mode.
All the transitions used in photon generation belong to the D2 line at around \SI{780}{\nano\meter}.
The total interrogation time is \SI{200}{\milli\second}.
The total cycle lasts $\sim \SI{1.25}{\second}$.

Due to their very different duty cycles, the two experiments run asynchronously.
The overall duty cycle of the experiment is limited to $\SI{5}{\percent}$.

\subsection{Laser locking setup}

Reducing the linewidths and maintaining resonance conditions by laser locking is important in our experiment.
It is crucial to maintain efficiency of the EIT storage in the Rydberg medium, which in terms of laser frequencies can be affected by drifting frequency of both the DLCZ read photon and the Rydberg coupling field.
The DLCZ read photon frequency is determined by the read pulse frequency (see Fig.~1 of the main text) which is locked to a saturated absorption spectroscopy signal offering stability on the order of \SI{200}{\kilo\hertz}, well below the EIT bandwidth $\sim \SI{2.3}{\mega\hertz}$.
To lock the coupling field frequency, we use a cascaded lock of two main elements: modulation transfer spectroscopy (MTS) \cite{Camy1982} and a transfer cavity.
We lock our probe laser at \SI{780}{\nano\meter} to the MTS signal using a method described in \cite{Escobar2015}.
Careful alignment of the MTS setup allowed us to reduce frequency drifts to below \SI{100}{\kilo\hertz} (limited by residual amplitude modulation introduced by an acousto-optic modulator).
A portion of the same laser light is sent to a \SI{25}{\cm}-long plano-concave transfer cavity of finesse $\sim 2000$ whose length is controlled by a piezo actuator to which one of the mirrors is glued.
The cavity length is stabilized with the standard Pound–Drever–Hall (PDH) technique \cite{Drever1983}.
A small portion of the coupling laser seed light at \SI{958}{\nm} is sent to the same transfer cavity and its frequency is also locked with PDH technique.
This allows for not only stabilization of the central frequency, but also to narrow the laser linewidth to $\sim \SI{60}{\kilo\hertz}$, improving the EIT transparency.
The linewidth was estimated by comparing the root-mean-square of the error signal and its slope.
Overall, the locking system allows us to perform experiments where the EIT storage efficiency remains constant for hours.

\subsection{Noise in the detection}

\begin{figure}
    \centering
    \includegraphics[scale=1.1]{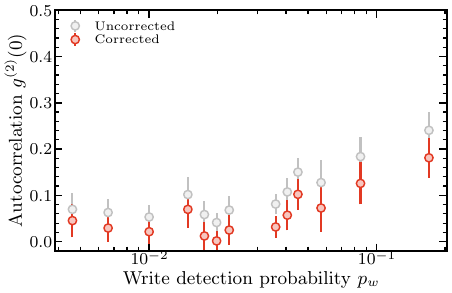}
    \caption{Noise correction on the autocorrelation measurements. We compare the uncorrected (grey circles) and corrected (red circles) autocorrelation functions at zero delay, as explained in the main text.}
    \label{G2corrvsuncorr}
\end{figure}

All the values of the autocorrelation function $g^{(2)}(0)$ reported in this paper are corrected for noise, which we assume is uncorrelated with the photon detections. 
It has two origins: i) dark counts of the detectors\footnote{In our experiment we used Laser Components COUNT-10C-FC and Excelitas SPCM-AQRH-14-FC} and ii) photon detections of the coupling field, and can be easily estimated from the detector clicks histograms in the region where the coupling beam is still on, but after the retrieved photonic pulse (see the schematic histogram in the Fig. 1 of the main text).
The noise has rates for detector D1 $\sim \SI{50}{\per\second}$ and detector D2 $\sim \SI{300}{\per\second}$, what corresponds to around \num{1.5e-5} and \num{9e-5} noise clicks per trial, respectively.

If one assumes that the noise clicks are uncorrelated with the ``good" photon clicks, one can show that the corrected $g^{(2)}(0)$ can be computed as
\begin{align}
    g^{(2)}(0) = g^{(2)}_n(0) - & \left(1 - g^{(2)}_n(0)\right) \nonumber \\
                & \times \left( \frac{n_{n, 1}}{n_1 - n_{n, 1}} + \frac{n_{n, 2}}{n_2 - n_{n, 2}} \right. \\
                & \quad + \left. \frac{n_{n, 1} n_{n, 2}}{(n_1 - n_{n, 1})(n_2 - n_{n, 2})} \right) \text{,} \nonumber
\end{align}
where $n_{n, 1(2)}$ and $n_{1(2)}$ are the mean number of noise clicks and detected signal clicks per trail, respectively, observed on detector 1(2) and $g^{(2)}_n(0)$ is the uncorrected (directly measured) autocorrelation function, computed as
\begin{equation}
    g^{(2)}_n(0) = \frac{n_{1,2} N}{n_1 n_2}
\end{equation}
with $n_{1,2}$ being the number of the detected coincidence clicks and $N$ the number of performed trials.
In \autoref{G2corrvsuncorr} one can see the effect of the correction on autocorrelation of the read photons after their storage in the non-linear medium.

\section{Theoretical description of the heralded photon source}

To understand the photon emission from the DLCZ photon source, we follow the theoretical model presented in \cite{Distante2017}.
We start from the ideal description of the DLCZ process and then include the noise contributions arising from spontaneous emission, background light fields and dark-counts of detectors.
We can write the photon detection probabilities as:
\begin{equation}
    \label{eqmodel}
    \begin{cases}
      p_{w} = p t_{w} + p_{nw}\\
      p_{r} = p \eta_{a} t_{r} + p(1-\eta_{a})p_{eg}t_{r} + p_{nr}\\
      p_{w,r} =  p_{w} \eta_{a} t_{r} + p_{w}p(1-\eta_{a})p_{eg}t_{r} + p_{w}p_{nr}
    \end{cases}\,.
\end{equation}
Here, $p_{w}$ is the write detection probability, $p$ is the probability that at least one excitation is generated in the cloud (in the desired write mode), $t_{w}$ is the transmission including the detector efficiency in the write path and $p_{nw}$ is the dark count probability.
Similarly, $p_{r}$ is the detection probability in the read path (unconditional), $t_{r}$ is the transmission including the detector efficiency in the read path, $p_{nr}$ is the dark count probability including background counts, $\eta_{a}$ is the intrinsic DLCZ read-out efficiency and $p_{eg}$ is the branching ratio $\ket{e_1} \rightarrow \ket{g_1}$ transition.
Finally, $p_{w,r}$ is the probability of having a coincidence click in the write and read mode in one experiment trial.
Note that each probability is a sum of a directional emission due to DLCZ and a noise contribution.
We ignored the dead time of detectors, considering that we experimentally never reach a regime where this becomes significant.

\begin{figure}[t]
    \begin{minipage}{\linewidth}
        \centering
        \includegraphics[width=\textwidth]{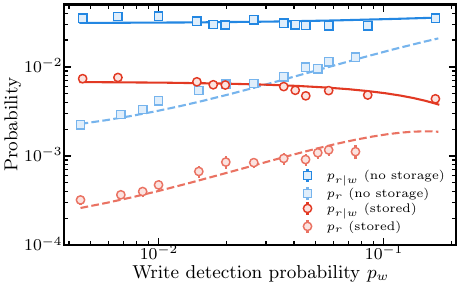}
    \end{minipage}
    \begin{minipage}{\linewidth}
        \centering
        \includegraphics[width=\textwidth]{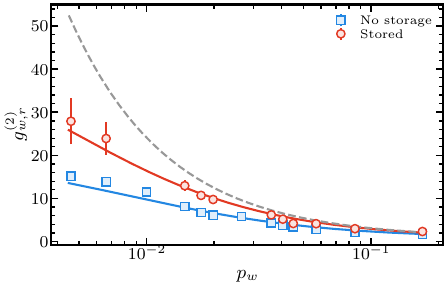}
    \end{minipage}
    \caption{DLCZ photon emission. 
    (upper): Conditional detection probabilities $p_{r|w}$ and unconditional probabilities $p_{r}$ are measured as a function of detected $p_{w}$ with and without storage in the Rydberg cloud. 
    Solid and dashed lines are the model predictions from \autoref{eqmodel} (see main text). 
    (lower): Cross-correlation $g^{(2)}_{w,r}(0)$ between write and read photons as a function of $p_{w}$, with and without storage. 
    Solid lines are the model prediction, in accordance with measurements. 
    The dashed line is the same as the solid lines, but the noise contributions $p_{nr}$ set to zero. 
    It collapses to the same value for both stored and not stored cases, which explains that the improvement of $g^{(2)}_{w,r}(0)$ is due to noise reduction.}
    \label{figg12}
\end{figure}

Inputting experimental parameters to this set of equation allows us to understand the measured probabilities.
The only free parameter is $p_{eg}$ which is difficult to measure or predict - the distribution of Zeeman states changes during the interrogation time, which affects the branching ratio.
Experimentally, we measure $\{ t_{w} = 0.21 \text{, } t_{r} = 0.09 \text{, } \eta_{a} = 0.32 \text{, } p_{nw} = 10^{-4} \text{, } p_{nr} = 1.5\cdot 10^{-3}\}$ and find $p_{eg} = 0.20$.
Measurements are presented alongside the model predictions in \autoref{figg12}(upper), with $p_{r|w}=p_{w,r}/p_{w}$ being the probability of detecting an event in the read mode conditioned on a prior write detection.
To include the effect of the Rydberg medium on DLCZ read photons, the transmission in the read mode is modified to $t_{r}(p_{w}) = \eta(p_{w})\cdot 0.09 $, where $\eta(p_{w})$ is the Rydberg memory storage efficiency which depends on $p_{w}$ (see Fig. 2.(inset) of the main text) and the numerical factor accounts for additional losses between the setups. Here, $\eta(p_{w})$ is directly taken from our measurements.
The read photon background probability is reduced to $p_{nr} = 1.3\cdot 10^{-4}$ thanks to the temporal and frequency filtering effect of the storage, as the read noise cannot be stored in the memory.

From this set of equations, we compute the cross-correlation function between the write and the read mode as
\begin{equation} \label{eq:g2}
    g^{(2)}_{w,r}= \frac{\langle a^{\dagger}_w a^{\dagger}_r a_r a_w \rangle}{\langle a^{\dagger}_w a_w \rangle \langle a^{\dagger}_r a_r\rangle} \approx \frac{p_{w,r}}{p_{w}p_{r}}\text{,}
\end{equation}
where $a^{\dagger}$ ($a$) is the creation (annihilation) operator of the write/read mode.
This quantity is often regarded as a quality metric for probabilistic single photon sources.
It is plotted  in \autoref{figg12}(lower), comparing the case when the read photons are stored in the Rydberg memory with the case of direct detection without storage.
We observe that the cross-correlation improves after the storage in the Rydberg memory.
However, we understand that this increase of $g^{(2)}_{w,r}(0)$ is due to a reduction of noise by temporal and frequency filtering in the EIT storage process.
Indeed, when re-plotting the two model predictions (green and blue solid lines) setting the noise contributions to zero, the noise-corrected curves collapse to the same value (gray dashed line).

To gain insight, we calculate the expected value of the cross correlation from the photon distribution in the ideal case, without noise or losses and considering a perfect retrieval efficiency.
We start by considering that the DLCZ produces a two-mode squeezed states for the write and read photons.
\begin{equation}\label{eq:twomodesqueezedstate}
    \ket{\Psi_{w,r}} = \sqrt{1-p}\sum\limits_{n=0}^{\infty}p^{n/2}\ket{n_{w},n_{r}}\text{,}
\end{equation}
with $p$ being the probability that at least one excitation is generated.
Calculating $g^{(2)}_{w,r}(0)$ according to \autoref{eq:g2} yields
\begin{equation}
    g^{(2)}_{w,r}(0) = 1 + \frac{1}{p}.
\end{equation}
In the ideal case of absence of transmission losses and that the non-linear medium is fully blockaded (hence, only one read photon is stored)  \autoref{eq:twomodesqueezedstate} needs to be re-written as
\begin{equation}\label{eq:rydberg_twomodesqueezedstate}
    \ket{\Psi_{w,r}} = \sqrt{1-p} \left( \ket{0_{w},0_{r}} + \sum\limits_{n=1}^{\infty}p^{n/2}\ket{n_{w},1_{r}} \right) \text{,}
\end{equation}
where the Fock component of the read mode is set to one.
Now, we expect to measure a slightly different cross-correlation of
\begin{equation}
    g^{(2)}_{w,r}(0) = \frac{1}{p}.
\end{equation}
One can notice that even if the system behaved perfectly, one would not expect any significant change on the cross-correlation $g^{(2)}_{w,r}(0)$.
In any case, we would not be able to discern such a small difference with our measurements.

\section{Estimation of multiphoton strength $\zeta$}

Let us consider an arbitrary state $\hat{\rho}_{\mathrm{in}}$ describing a photonic pulse.
In case of our experiment, it can be the state of the DLCZ read photon when leaving the DLCZ cloud.
Its Fock state distribution can be represented as a vector with elements defined as
\begin{equation}
    \label{eq:fock_distribution}
    p_k = \bra{k} \hat{\rho}_{\mathrm{in}} \ket{k}\text{,}
\end{equation}
where $\ket{k}$ is $k$-photon Fock state.

\subsection{Transfer matrix formalism}

Any element affecting this distribution, let say a beam splitter or transmission losses, can be described by a matrix $M$, such that the Fock state distribution after this element can be expressed as
\begin{equation}
    p'_k = M_{kl} p_l ,
\end{equation}
where $M_{kl}$ are elements of the matrix and the summation over the repeating indices is implicit.
For this matrix to preserve the normalization of the Fock state distribution, i.e. $\sum_k p'_k = 1$, it's necessary that its columns sum up to 1. 

Virtually any element can be expressed as such a matrix.
A typical example would be a beam splitter or any other lossy element with transmission $t$.
If we aim to determine the Fock state distribution after this element, the matrix elements for this operation can be described by the binomial probability mass function:
\begin{equation}
    \label{eq:binomial}
    M_{kl} = \binom{l}{k} t^{k} (1 - t)^{l - k} \text{.}
\end{equation}
Consider another example: a perfect single-photon filter, such as a Rydberg ensemble that's both perfectly blockaded and perfectly transmissive.
This filter transforms all higher order Fock components into 1-photon component.
The matrix representation for such a filter would be:
\begin{equation}
    M = 
    \begin{pmatrix}
    1 & 0 & 0 & 0 & \cdots \\
    0 & 1 & 1 & 1 & \cdots \\
    0 & 0 & 0 & 0 & \cdots \\
    \vdots & \vdots & \vdots & \vdots & \ddots \\
    \end{pmatrix}.
\end{equation}

Generally, lossy operations are represented by upper triangular matrices. 
However, if one were to model elements adding photons, such as detector dark counts, using this approach, the matrix would exclusively have elements in the lower triangle.
A nice bonus of this formalism is how straightforward it is to back-propagate any operation just by inverting its matrix representation (provided that the matrix is not singular).

Knowing how we can easily account for transmission losses in our system, we can estimate what is the Fock state distribution at the input of the Rydberg cloud.

\subsection{Coherent state}

Determining the Fock state distribution for a coherent state is a straightforward task because it follows the Poisson distribution.
Moreover, any linear element retains this characteristic.
Therefore, for a coherent state, the Fock state distribution is expressed as
\begin{equation}
    \label{eq:fock_wcs}
    p_k = e^{-|\alpha|^2} \frac{|\alpha|^{2k}}{k!}\text{,}
\end{equation}
where $|\alpha|^2 = \mu_{\mathrm{in}}$, with $\mu_{\mathrm{in}}$ being the mean input photon number, derived from the back-propagation of loss from the detection probabilities at single-photon detectors.
Finally, one obtains multiphoton strength $\zeta$ using the expression from the main text.

\subsection{DLCZ single photons}

Determining the Fock state distribution for the DLCZ read photon at the Rydberg ensemble's input is more complex.
We start by considering that the DLCZ produces a two-mode squeezed state for the write and read photons
\begin{equation}
    \label{eq:two_mode_squeezed_state}
    \ket{\Psi_{w,r}} = \sqrt{1-p} \sum\limits_{n=0}^{\infty} p^{n/2} \ket{n_{w},n_{r}} \text{,}
\end{equation}
with $p$ being the probability that at least one excitation is generated.
We are interested in the Fock state distribution in the read mode, conditioned on a prior detection in the write mode.
We model the detection of a write photon by the following POVM operator, which takes into account transmission and detection efficiencies ($t_{w}$) and models non-photon-number-resolving detection:
\begin{equation}
    \hat{\Pi}_{\mathrm{det}} = \sum\limits_{n=1}^{\infty}\left[ 1 - \left(  1 - t_{w} \right)^{n}  \right]\ket{n_{w}}\bra{n_{w}}.
\end{equation}
The resulting conditional density matrix for the read-mode $\hat{\rho}_{r|w}$ can be written as: 
\begin{align}
    \label{eq:dlcz_read_state}
    \hat{\rho}_{r|w} = & \frac{ \text{Tr}_{w} \left[ \hat{\Pi}_{\mathrm{det}} \hat{\rho}_{w,r} \right] }{\text{Tr}_{w,r} \left[ \hat{\Pi}_{\mathrm{det}} \hat{\rho}_{w,r}  \right]} \nonumber \\
    = & (1 - p) t^{-1}_{w} \left[ 1 - p (1 - t_{w}) \right] \nonumber \\
    & \times \sum\limits_{n=1}^{\infty }p^{n-1} \left[ 1 - \left( 1 - t_{w} \right)^{n} \right] \ket{n_{r}} \bra{n_{r}} \text{,}
\end{align}
where $\hat{\rho}_{w,r} = \ket{\Psi_{w,r}}\bra{\Psi_{w,r}}$.
We can identify $\hat{\rho}_{w,r}$ with $\hat{\rho}_{\mathrm{in}}$ from \autoref{eq:fock_distribution} and use the above-described transfer matrix formalism to account for the losses between the setups $t_{r} = T \eta_{a} = 0.15$, with $T$ being the transmission factor between the two experiments and $\eta_{a}$ being the DLCZ read-out efficiency.
The remaining task to determine multiphoton strength is to figure out the value of $p$ in \autoref{eq:two_mode_squeezed_state}.

Due to losses between the generation of DLCZ photons and their detection, it would be difficult to directly measure their Fock state distribution and infer $p$ \footnote{
Alternatively, one could infer the generated two-mode squeezed state \autoref{eq:two_mode_squeezed_state} based on the measured $p_w$, but this would require very careful calibration of losses in the write path, which, from our experience, can be challenging.
}.
Instead, we estimate it from the auto-correlation function $g^{(2)}(0)$, which is independent of linear losses and is bijective for given losses $t_r$.
To obtain $p$, we find a probability distribution $p_k = \bra{k} \hat{\rho}_{r|w} \ket{k}$ that yields given $g^{(2)}(0)$.
A general formula for the auto-correlation function $g^{(2)}(0)$ of an arbitrary state is
\begin{equation}
    g^{(2)}(0) = \frac{\langle a^{\dagger}a^{\dagger}a a \rangle}{\langle a^{\dagger}a\rangle^2} = \frac{\sum\limits_{k=0}^{\infty} k (k - 1) p_k}{\left[ \sum\limits_{k=0}^{\infty} k p_k \right]^{2}}\text{.}
\end{equation}
Then, from $p$, we can infer the input Fock state distribution and the corresponding multiphoton strength $\zeta$.

As a side note, we mention here that we also observe that the measured $g^{(2)}(0)$ of the light generated with the DLCZ protocol does not obey the expected scaling $2p (2 + p) / (1 + p)^{2}$ \cite{Distante2017}. 
However, we were not able to find an explanation to this result.

\section{Monte Carlo simulation of the partial blockade}

In our numerical simulation, we simplify the problem to one dimension.
This choice is motivated by the directionality of polariton propagation and the small transverse extent of the probe mode.
We also assume the ensemble has a uniform density distribution over its length, determined by the cloud's FWHM.
Additionally, we model the dipole blockade as a binary effect: if two polaritons are closer than the blockade radius, the later-arriving one gets scattered; otherwise, both propagate without losses.

It's important to note that this is an effective model, and some physical aspects of the problem are not addressed.
Specifically, the binary blockade is a good approximation only in ensembles with a very high optical depth \cite{Zeuthen2017, Bienias2020}.
However, by storing the information, we amplify the nonlinearity \cite{Distante2016} -- any photon not stored as a polariton, even if unscattered, is essentially filtered in time by the storage process.
We also overlook the dephasing interactions between polaritons during storage \cite{Busche2017}, which can be viewed as another nonlinearity booster.
Both of these factors might be the reason why the model's blockade radius matches so closely with what's expected from the usual formula for blockade radius \cite{Firstenberg2016}.
It would be worthwhile to vary the storage time and see if this affects the determined blockade radius.
Additionally, we do not account for more nuanced effects that especially arise at high multiphoton strengths, such as dissipative interactions \cite{Zeuthen2017} and pollutants \cite{Bienias2020}.
Both are likely to elevate the measured $g^{(2)}(0)$ by diminishing the $g^{(2)}(0)$ expression's denominator.
Yet, we observe pollutant effects only when $\zeta$ is near 1.
Overall, our simulation should be seen as a tool that offers an intuition of how the medium modifies the Fock state distribution of stored pulses.

In our simulation, we start by randomly selecting a number of positions within the cloud corresponding to the number of a  specific input Fock state.
Sequentially analyzing these positions, we determine the feasibility of polariton presence at each spot by considering the blockade effect.
To clarify, there may be scenarios where the third polariton survives without being lost, even if it's located closer than the blockade radius to the second one.
This can happen if the second polariton was previously scattered by the first.
We repeat this process \num{1e5} times for each Fock state to determine the probability distribution of survival and conversion into other Fock states.
The derived distributions are then integrated as columns in a transfer matrix, capturing the influence of (partial) blockade on the input Fock state distribution.

To determine the Fock state distribution after the storage in the Rydberg ensemble, we utilize the transfer matrix formalism described earlier.
For DLCZ photons, we begin with \autoref{eq:dlcz_read_state} as the input state and compute its Fock state distribution.
Subsequently, this is modified by matrices that factor in transmission losses between setups ($t_\mathrm{losses} = 0.15$), the fact that the input pulse cannot be fully compressed within the medium (resulting in a part only of the pulse being effectively stored which we model as a beam slitting operation, $\eta_\mathrm{compression} = 0.6$), and half the linear losses during EIT propagation ($\sqrt{\eta_\mathrm{EIT}} = \sqrt{0.6}$) \footnote{
The remaining linear losses during EIT propagation do not affect the measured $g^{(2)}(0)$, however, they need to be included in the calculation of the storage and retrieval efficiency, as it is shown in the next section.
}.
Estimating the second factor is somewhat challenging due to complex propagation dynamics of such short pulses under rEIT conditions \cite{Mohl2020, Padron-Brito2021a}.
Nonetheless, the precise value influences our simulation results only weakly, evident in fig. 3 of the main text where the limits of the shaded area correspond to $\eta_\mathrm{compression}$ being 0.45 and 0.75.
Finally, the matrix derived from the Monte Carlo simulation is used to account for the blockade effect.

This procedure can be concisely represented by:
\begin{equation}
    \label{eq:distr_after_blockade}
    p' = M^\mathrm{blockade} \, M^{\sqrt{\mathrm{EIT}}} \, M^\mathrm{compression} \, M^\mathrm{losses} \, p \ ,
\end{equation}
where elements of $p$ are $p_k =  \bra{k} \hat{\rho}_{r|w} \ket{k}$ with $\hat{\rho}_{r|w}$ as per \autoref{eq:dlcz_read_state}.
The matrices, $M$, are defined as outlined above.

In the case of coherent states, we don't need to account for the transmission losses, because by backpropagation we calculate directly the state right before the cloud.

This simulation can also be employed to mimic results from slow-light propagation without storage.
To adjust for pulses that cannot be entirely compressed within the medium, one might consider expanding the medium size to align with the pulse length.
Yet, to align with the experimentally observed results, when using the blockade radius extracted from storage data, the medium needs to be more than twice as large as above reasoning would indicate.
This discrepancy is most probably caused by two factors.
Firstly, unlike stored pulses, propagating pulses don't benefit from the two nonlinearity-enhancing effects mentioned earlier in this section.
Consequently, the nonlinearity they experience is weaker.
Secondly, previous observations have indicated that the initial segments of pulses undergoing rEIT propagation exhibit minimal reduction in $g^{(2)}(0)$ \cite{Mohl2020, Padron-Brito2021a}.

Despite their simplicity, our simulations align well with our data and provide a clear insight into how the partial blockade impacts the Fock state distribution of the incoming pulses.

\section{Storage and retrieval efficiency}

Once the Fock state distribution of the pulses post-blockade is determined, simulating the storage and retrieval efficiency is straightforward.
One needs to account for two factors, the remaining linear losses during EIT propagation ($\sqrt{\eta_\mathrm{EIT}} = \sqrt{0.6}$) and the retrieval efficiency ($\eta_r = 0.41$).
This efficiency is deduced by matching the simulated efficiency with the experimental results at very low multiphoton strengths, where the influence of the blockade is negligible.

Given these considerations, the efficiency for the DLCZ photons can be calculated as:
\begin{equation}
    \eta = \frac{\mu_s}{\mu_{\mathrm{in}}}
    = \frac{ \eta_{r} \sum\limits_{k=0}^{\infty} k M^{\sqrt{\mathrm{EIT}}}_{kl} p'_l} {\sum\limits_{k=0}^{\infty} k M^\mathrm{losses}_{kl} \, p_l }
    = \frac{ \eta_{r} \sqrt{\eta_\mathrm{EIT}} \sum\limits_{k=0}^{\infty} k \, p'_k} {t_\mathrm{losses} \sum\limits_{k=0}^{\infty} k \, p_k }, 
\end{equation}
where the last equality comes from the linearity of the losses.
Here, $\mu_{\mathrm{in}}$ and $\mu_s$ denote the mean input and retrieved photon numbers, respectively.
$p_l$ and $p'_l$ represent the elements of Fock state distributions as in \autoref{eq:distr_after_blockade}.

Finally, we use the same method to estimate the storage and retrieval efficiency for coherent states.

\section{DLCZ photons vs. coherent states distribution}

\begin{figure}[t]
    \centering
    \includegraphics[width=\columnwidth]{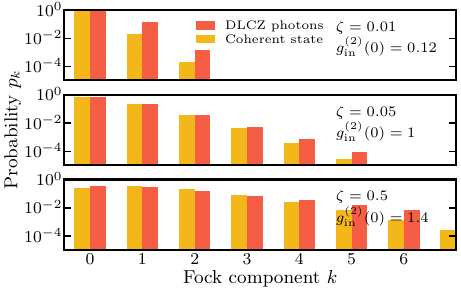}
    \caption{Input Fock state distribution of DLCZ read photon and coherent state for the same values of $\zeta$. We plot it for $\zeta = \{ 0.01,0.05,0.5\}$, corresponding to a $g^{2}_{\mathrm{in}}(0) = \{ 0.12, 1, 1.4\}$ of the DLCZ generated light field. Coherent states have always a $g^{2}_{\mathrm{in}}(0) = 1$ whatever the $\zeta$ value.
    The transmission losses are included in the computation of the Fock state distributions.
    }
    \label{fig:fock_dist_dlcz_vs_wcs}
\end{figure}

In the main text, we mentioned that the distinct trends observed between the DLCZ photons and coherent states -- particularly regarding efficiency and stored $g^{(2)}(0)$ at high multiphoton strengths -- stem from the differences in their Fock state distributions.
This difference becomes evident in \autoref{fig:fock_dist_dlcz_vs_wcs}.
As $\zeta$ values rise, the DLCZ Fock state distribution exhibits a more pronounced "tail."
This leads to a more marked reduction in storage efficiency and a faster rise in $g^{(2)}(0)$ after its interaction with the non-linear medium.

\bibliography{purification}

%% file: figures/Setup_Purification.pdf_tex
\begingroup%
  \makeatletter%
  \providecommand\color[2][]{%
    \errmessage{(Inkscape) Color is used for the text in Inkscape, but the package 'color.sty' is not loaded}%
    \renewcommand\color[2][]{}%
  }%
  \providecommand\transparent[1]{%
    \errmessage{(Inkscape) Transparency is used (non-zero) for the text in Inkscape, but the package 'transparent.sty' is not loaded}%
    \renewcommand\transparent[1]{}%
  }%
  \providecommand\rotatebox[2]{#2}%
  \newcommand*\fsize{\dimexpr\f@size pt\relax}%
  \newcommand*\lineheight[1]{\fontsize{\fsize}{#1\fsize}\selectfont}%
  \ifx\svgwidth\undefined%
    \setlength{\unitlength}{452.82301054bp}%
    \ifx\svgscale\undefined%
      \relax%
    \else%
      \setlength{\unitlength}{\unitlength * \real{\svgscale}}%
    \fi%
  \else%
    \setlength{\unitlength}{\svgwidth}%
  \fi%
  \global\let\svgwidth\undefined%
  \global\let\svgscale\undefined%
  \makeatother%
  \begin{picture}(1,0.41699125)%
    \lineheight{1}%
    \setlength\tabcolsep{0pt}%
    \put(0,0){\includegraphics[width=\unitlength,page=1]{Setup_Purification.pdf}}%
    \put(0.15430082,0.16721469){\color[rgb]{0,0,0}\rotatebox{-0.42224489}{\makebox(0,0)[lt]{\lineheight{1.25}\smash{\begin{tabular}[t]{l}$\{$\end{tabular}}}}}%
    \put(0.74247726,0.18016461){\color[rgb]{0,0,0}\rotatebox{-0.36067414}{\makebox(0,0)[lt]{\lineheight{1.25}\smash{\begin{tabular}[t]{l}$\ket{r}$ \end{tabular}}}}}%
    \put(0.74268431,0.09155382){\color[rgb]{0,0,0}\rotatebox{-0.36067414}{\makebox(0,0)[lt]{\lineheight{1.25}\smash{\begin{tabular}[t]{l}$\ket{e_{2}}$ \end{tabular}}}}}%
    \put(0.7426842,0.03192816){\color[rgb]{0,0,0}\rotatebox{-0.36067414}{\makebox(0,0)[lt]{\lineheight{1.25}\smash{\begin{tabular}[t]{l}$\ket{g_{2}}$\end{tabular}}}}}%
    \put(0.37167698,0.03192928){\color[rgb]{0,0,0}\rotatebox{-0.36067414}{\makebox(0,0)[lt]{\lineheight{1.25}\smash{\begin{tabular}[t]{l}$\ket{s}$\end{tabular}}}}}%
    \put(0.37167706,0.07167974){\color[rgb]{0,0,0}\rotatebox{-0.36067414}{\makebox(0,0)[lt]{\lineheight{1.25}\smash{\begin{tabular}[t]{l}$\ket{g_{1}}$\end{tabular}}}}}%
    \put(0.37167728,0.17651049){\color[rgb]{0,0,0}\rotatebox{-0.36067414}{\makebox(0,0)[lt]{\lineheight{1.25}\smash{\begin{tabular}[t]{l}$\ket{e_{1}}$\end{tabular}}}}}%
    \put(0.24580025,0.03192961){\color[rgb]{0,0,0}\rotatebox{-0.36067414}{\makebox(0,0)[lt]{\lineheight{1.25}\smash{\begin{tabular}[t]{l}$\ket{s}$\end{tabular}}}}}%
    \put(0.24580032,0.07168007){\color[rgb]{0,0,0}\rotatebox{-0.36067414}{\makebox(0,0)[lt]{\lineheight{1.25}\smash{\begin{tabular}[t]{l}$\ket{g_{1}}$\end{tabular}}}}}%
    \put(0.24580054,0.17651018){\color[rgb]{0,0,0}\rotatebox{-0.36067414}{\makebox(0,0)[lt]{\lineheight{1.25}\smash{\begin{tabular}[t]{l}$\ket{e_{1}}$\end{tabular}}}}}%
    \put(0,0){\includegraphics[width=\unitlength,page=2]{Setup_Purification.pdf}}%
    \put(0.13848279,0.34738757){\color[rgb]{0,0,0}\makebox(0,0)[t]{\lineheight{1.25}\smash{\begin{tabular}[t]{c}3\end{tabular}}}}%
    \put(0.3575214,0.22800408){\color[rgb]{0,0,0}\makebox(0,0)[t]{\lineheight{1.25}\smash{\begin{tabular}[t]{c}1\end{tabular}}}}%
    \put(0.16553703,0.25577894){\color[rgb]{0,0,0}\makebox(0,0)[t]{\lineheight{1.25}\smash{\begin{tabular}[t]{c}2\end{tabular}}}}%
    \put(0.33513143,0.29943307){\color[rgb]{0,0,0}\makebox(0,0)[t]{\lineheight{1.25}\smash{\begin{tabular}[t]{c}4\end{tabular}}}}%
    \put(0.07103343,0.13620417){\color[rgb]{0,0,0}\makebox(0,0)[t]{\lineheight{1.25}\smash{\begin{tabular}[t]{c}D1\end{tabular}}}}%
    \put(0.84269795,0.16782131){\color[rgb]{0.01176471,0.00784314,0}\makebox(0,0)[t]{\lineheight{1.25}\smash{\begin{tabular}[t]{c}D2\end{tabular}}}}%
    \put(0.93526344,0.07093692){\color[rgb]{0.01176471,0.00784314,0}\makebox(0,0)[t]{\lineheight{1.25}\smash{\begin{tabular}[t]{c}D3\end{tabular}}}}%
    \put(0.19337692,0.18678057){\color[rgb]{0,0,0}\makebox(0,0)[t]{\lineheight{1.25}\smash{\begin{tabular}[t]{c}1\end{tabular}}}}%
    \put(0.20985231,0.18691695){\color[rgb]{0,0,0}\makebox(0,0)[t]{\lineheight{1.25}\smash{\begin{tabular}[t]{c}2\end{tabular}}}}%
    \put(0.32024095,0.18661592){\color[rgb]{0,0,0}\makebox(0,0)[t]{\lineheight{1.25}\smash{\begin{tabular}[t]{c}3\end{tabular}}}}%
    \put(0.33681782,0.18622243){\color[rgb]{0,0,0}\makebox(0,0)[t]{\lineheight{1.25}\smash{\begin{tabular}[t]{c}4\end{tabular}}}}%
    \put(0.22083983,0.38113659){\color[rgb]{0,0,0}\makebox(0,0)[t]{\lineheight{1.25}\smash{\begin{tabular}[t]{c}Heralded non-classical light source\end{tabular}}}}%
    \put(0.77944857,0.37888401){\color[rgb]{0,0,0}\makebox(0,0)[t]{\lineheight{1.25}\smash{\begin{tabular}[t]{c}Non-linear Rydberg quantum memory\end{tabular}}}}%
    \put(0.49638079,0.26657373){\color[rgb]{0,0,0}\makebox(0,0)[t]{\lineheight{1.25}\smash{\begin{tabular}[t]{c}$g^{(2)}_{\mathrm{in}}(0)$\end{tabular}}}}%
    \put(0,0){\includegraphics[width=\unitlength,page=3]{Setup_Purification.pdf}}%
    \put(0.60973371,0.19676291){\color[rgb]{0,0,0}\makebox(0,0)[t]{\lineheight{1.25}\smash{\begin{tabular}[t]{c}5\end{tabular}}}}%
    \put(0.69786396,0.19676291){\color[rgb]{0,0,0}\makebox(0,0)[t]{\lineheight{1.25}\smash{\begin{tabular}[t]{c}6\end{tabular}}}}%
    \put(0,0){\includegraphics[width=\unitlength,page=4]{Setup_Purification.pdf}}%
    \put(0.67681269,0.3002802){\color[rgb]{0.00392157,0.00392157,0}\makebox(0,0)[t]{\lineheight{1.25}\smash{\begin{tabular}[t]{c}5\end{tabular}}}}%
    \put(0.85521,0.3002802){\color[rgb]{0,0,0}\makebox(0,0)[t]{\lineheight{1.25}\smash{\begin{tabular}[t]{c}6\end{tabular}}}}%
    \put(0,0){\includegraphics[width=\unitlength,page=5]{Setup_Purification.pdf}}%
    \put(0.78243732,0.32315751){\color[rgb]{0.44313725,0.13333333,0.48627451}\makebox(0,0)[t]{\lineheight{1.25}\smash{\begin{tabular}[t]{c}$r_{b}$\end{tabular}}}}%
    \put(0.11409666,0.20576784){\color[rgb]{0,0,0}\makebox(0,0)[t]{\lineheight{1.25}\smash{\begin{tabular}[t]{c}FPC\end{tabular}}}}%
    \put(0.94816578,0.20007547){\color[rgb]{0,0,0}\makebox(0,0)[t]{\lineheight{1.25}\smash{\begin{tabular}[t]{c}7\end{tabular}}}}%
    \put(0.66791026,0.25763108){\color[rgb]{0,0,0}\makebox(0,0)[t]{\lineheight{1.25}\smash{\begin{tabular}[t]{c}DM\end{tabular}}}}%
    \put(0.86334966,0.25763108){\color[rgb]{0,0,0}\makebox(0,0)[t]{\lineheight{1.25}\smash{\begin{tabular}[t]{c}DM\end{tabular}}}}%
    \put(0.95196047,0.15784044){\color[rgb]{0,0,0}\makebox(0,0)[t]{\lineheight{1.25}\smash{\begin{tabular}[t]{c}BS\end{tabular}}}}%
    \put(0.14509262,0.16667233){\color[rgb]{0,0,0}\makebox(0,0)[t]{\lineheight{1.25}\smash{\begin{tabular}[t]{c}$\Delta$\end{tabular}}}}%
    \put(0.88223485,0.0358544){\color[rgb]{0,0,0}\makebox(0,0)[t]{\lineheight{1.25}\smash{\begin{tabular}[t]{c}$g^{(2)}_{\mathrm{out}}(0)$\end{tabular}}}}%
    \put(0.24409587,0.24554156){\color[rgb]{0,0,0}\makebox(0,0)[t]{\lineheight{1.25}\smash{\begin{tabular}[t]{c}E1\end{tabular}}}}%
    \put(0.7674795,0.24554156){\color[rgb]{0,0,0}\makebox(0,0)[t]{\lineheight{1.25}\smash{\begin{tabular}[t]{c}E2\end{tabular}}}}%
  \end{picture}%
\endgroup%